# Heliyon

# Development of routing algorithms in networks-on-chip based on ring circulant topologies

Aleksandr Yu. Romanov[*]

*National Research University Higher School of Economics, 34 Tallinskaya Ulitsa, Moscow, 123458, Russian Federation*

[*] Corresponding author.
E-mail address: a.romanov@hse.ru (A.Yu. Romanov).

## Abstract

This work is devoted to the study of communication subsystem of networks-on-chip (NoCs) development with an emphasis on their topologies. The main characteristics of NoC topologies and the routing problem in NoCs with various topologies are considered. It is proposed to use two-dimensional circulant topologies for NoC design, since they have significantly better characteristics than most common mesh and torus topologies, and, in contrast to many other approaches to improving topologies, have a regular structure. The emphasis is on using ring circulants which although in some cases have somewhat worse characteristics than the optimal circulants, compensate by one-length first generatrix in such graphs that greatly facilitate routing in them. The paper considers three different approaches to routing in NoCs with ring circulant topology: Table routing, Clockwise routing, and Adaptive routing. The algorithms of routing are proposed, the results of synthesis of routers, based on them, are presented, and the cost of chip resources for the implementation of such communication subsystems in NoCs is estimated.

Keyword: Electrical engineering

# 1. Introduction

New tasks, performed by systems-on-chip (SoC), require an increasing performance. Both their complexity and heterogeneity, as well as the area of SoC, increase. These lead to the fact that the need for more effective mechanisms of communication is constantly growing. Here, the experience of traditional computer networks, embodied in the paradigm of networks-on-chip (NoCs), can help. NoCs are designed to improve the following SoC parameters [1]: bandwidth scalability in comparison with traditional bus architectures; SoC energy efficiency and reliability; ready-made solutions for building different NoCs.

A typical NoC consists of a set of nodes, called IPs, which can be processor cores, built-in memory, separate specialized hardware modules, I/O interfaces. Each IP typically has one router to provide packet routing. Intermediate layer, which is a network interface, connects routers with IP nodes. All these components are interconnected by connecting lines [2].

The theory and practice of NoC development has been steadily increasing every year. There are a large number of different solutions at the level of synchronization of NoC communication subsystem [3, 4, 5], communication technology [5, 6, 7], connecting links organization [8], routers organization (packet-level connection technology [2, 9], route search algorithms [10], flow control [11]), and NoCs quality-of-service (QoS) arbitrage [2, 4, 9]. The analysis of this wide range of solutions shows that the most common NoCs are synchronous [3] or GALS networks [5] with packet communication, distributed generation of destination addresses, wormhole/virtual channels [12] with packet-level connection and the availability of QoS mechanisms [4]. This is because such networks provide acceptable bandwidth and well adapt to various tasks, and that's why there will be considered networks of this kind in this work. Moreover, NoCs are constantly evolving; new topologies and router schemes, as well as other solutions are being introduced. Since the main task of a NoC is to provide communication between IP nodes, it is obvious that network reliability depends on the use of an effective communication strategy between individual NoC nodes. Thus, development and implementation of new routing architectures ensuring reliable and fast transfer of packets between source and destination nodes is an important and urgent task.

# 2. Background

## 2.1. NoC topologies and routing in them

NoC effectiveness is decisively influenced by NoC topology, which greatly affects the structure of routers, routing algorithm, and costs of connectivity [13]. Choosing the optimal topology for a particular task can significantly improve NoC characteristics as a whole. In the general case, NoC topology is an undirected connected graph consisting of vertices — routers, and edges — physical links between them. The basic characteristics of NoC topology are as follows: number of vertices ($N$) — computational nodes; number of edges ($Ed$) — physical connections between routers; degree of a vertex ($St$) — number of edges emanating from it; graph diameter ($D$) — maximum among the minimum distances between any two vertices; average distance between nodes ($L_{av}$) — mean value of the shortest paths between all nodes of the graph; etc. [14, 15]. All these parameters significantly influence the efficiency of routing algorithms and networks connectivity topologies used in NoCs.

## 2.2. Problems that arise when routing in NoCs

Deadlocks occur when two or more packets wait to release each other's link or buffer as a result of existence of cycles in the network. To combat deadlock, there are several strategies. So, with deterministic routing, one can ensure the absence of cycles that will ensure the impossibility of deadlocks. Nevertheless, this is fraught with a greater danger of the influence of congestion and network errors on the bandwidth of NoC, since deterministic routing does not provide for the choice of alternative paths [10]. To prevent deadlocks, there is an approach based on virtual channels providing alternative routes for passing packets bypassing blocked channels first proposed in [16]. Another widespread approach was based on the method called "Turn Model" [17], which predicts deadlocks and changes packet direction, in such a way, eliminating cycles. Most modern algorithms also use adaptive routing which avoids most locks, but can lead to active deadlocks, such as "Hot potato" algorithm used in Nostrum NoC [18]. Active deadlocks occur when a packet cannot reach the destination address for a long time.

There are also congestions (most of the traffic, concentrated in certain places of the network, while other connections are idle) [19, 20] and network faults (can be permanent due to damaged connections and temporary problems) [21].

It should be noted that most problems arise when routing in NoCs are solved at the level of the router and the choice of correct packet routing strategy, while the cause of these problems is largely the selected NoC topology. Thus, the topological approach to the design of NoCs will potentially allow to partially reduce the impact of those routing problems that arise in NoCs.

## 2.3. Regular mesh-like NoC topologies

Regular topologies include all topologies with a scalable homogeneous structure most common of which are mesh topology and its improvements — torus and folded torus. In addition, mesh and torus can have both 2D and 3D shapes. According to [10], where 136 papers were analyzed, the most common in mesh-like NoCs is the XY routing algorithm. The peculiarity of the algorithm is its simplicity, because it uses the regularity of mesh topology and previously known geometric "form" of location of network nodes. This allows moving packets first, horizontally and then, vertically. Thus, routers do not need to store routing tables; it's enough to compare the address of the packet destination node with its own address, and then, via a simple algorithm, the packet is sent to the desired port. In the head pack flit, it is also sufficient to store only the destination address that allows reducing the amount of transmission of overhead in the network. Nevertheless, since the algorithm is deterministic, it is vulnerable to deadlocks, congestions, and network faults. This led to appearance of its numerous modifications by introducing adaptability and the use of virtual channels [16]. According to [10], with uniform traffic in mesh-like networks, deterministic algorithms in comparison with adaptive algorithms provide higher bandwidth. But even with unevenly distributed traffic, as well as using torus topology, adaptive algorithms outperform deterministic ones. Deterministic algorithms underutilize transmission channels, while adaptive algorithms distribute traffic more evenly in the network due to the presence of several routes for passing of packets. However, the router's circuit, that implements the adaptive routing algorithm, is more complex and also requires transfer of additional overhead that results in the consumption of additional resources.

Other regular topologies, such as hypercube (*n*-cube) [14, 15], *n*-dimensional cubically connected cycles (CCC) [16], WK-recursive topology [22], omega and butterfly topologies, Benes networks [23], and many other topologies have not found wide application in NoCs due to their worse characteristics and more complex routing algorithms. Often, there is no unambiguous routing algorithm in such networks that makes it necessary to resort to routing tables, or use complex nondeterministic routing algorithms.

### 2.4. Improved non-regular mesh-like NoC topologies

Mesh topology has a relatively low hardware cost and $Ed$, but a large $D$ and a correspondingly large $Lav$. Torus topology has maximum resource costs for standard routers with 5 ports (1 local and 4 external), but relatively small $D$ and $Lav$. However, a large step between topologies of different dimensions ($Ext$) is a significant drawback. Optimal are topologies with the number of nodes that are powers of natural numbers. With a different number of nodes, developers are forced to refer to rectangular topologies that are not optimal, or to use additional routers.

Some studies suggest ways to improve known regular topologies by implementing additional long irregular links to reduce the diameter of the topology. In [24] it is proposed to extend the mesh topology by 8 regular connections. The number of additional connections does not change with the increase in dimensionality of the topology; however, it makes it possible to halve the diameter. However, this approach requires using more bulky 6 port routers. In [25] an approach, involving reduction of critical load of mesh topology by implementing additional irregular long connections between nodes with large critical traffic between them, is proposed. This approach reduces the average distance between nodes by 13.1 % and critical load of the links. However, this leads to an increase in resource costs, as well as appearance of irregularities in the topology and, accordingly, an increase in deadlock probability which is solved using a specialized routing algorithm [26] based on the theory of eliminating cycles in graphs, or using routing tables.

Thus, use of different approaches to improving characteristics of mesh-like networks by completing additional connections leads firstly, to complexity and increase in dimension of routers, and secondly, to loss of regularity of topologies themselves and complication of routing algorithm.

### 2.5. Irregular topologies

A separate class is represented by irregular topologies that are used in case of development of specialized NoCs for the performance of a specific narrow problem when its characteristic graph and distribution of data flows between nodes are known in advance. Topologies, synthesized to optimize one or more network metrics, are also irregular. For example, the number of connections between nodes can be minimized if the goal is to reduce the hardware costs of connections; as well as network diameter, if it is necessary to ensure a guaranteed maximum packet delivery distance. With a large number of nodes, the task of finding a global optimum for a given objective function is too large for computational resources, and as a result, it is usually limited to the search of a "good" solution using such method as evolutionary computation. This approach is exemplified by quasi-optimal topologies [27]. Such topologies cannot be described in any way as a set of rules. Therefore, routing in them is possible only with the help of unique routing tables stored in each router or learning-based algorithms [20]. Sometimes path is calculated in IP nodes and is contained in the head flit. These approaches are resource-intensive that largely offset the benefits achieved by the optimizing topology [27].

## 3. Study area

### 3.1. Circulant topologies as an alternative to classical regular topologies

There is a separate class of regular topologies that were not considered together with mesh-like topologies. They are circulant topologies. To consider them, there is the need to give them a mathematical description: a circulant topology is a graph consisting of $N$ vertices and a set $S = \{1 \leq s_1 < \ldots < s_k < N\}$ of generatrices which can be described as $C(N; s_1, \ldots, s_k)$. Parameter $k$ specifies dimension of the graph and determines set of its edges $E = \{(v, v \pm mod(s_i, N)) \mid v \varepsilon V, i = \overline{1,k}\}$, the number of which is $2k \cdot N$.

Thus, circulant network $C(N; s_1, \ldots, s_k)$ can be represented as a ring-like structure, where each vertex is associated with $k$ successive vertices and $k$ previous vertices in steps of $s_1, \ldots, s_k$ (Fig. 1). Circulants with one-length first generatrix called ring circulants (Fig. 1a).

For implementation as NoC topologies, circulants of dimension 2 are of primary interest. This class of circulants has a degree of vertices 4 and good topological characteristics which makes it promising for use as a basis for NoCs. Since relatively small routers with 4 inputs/outputs are suitable for their construction, plenty of examples were developed (for example, Netmaker library [28]), and their high efficiency and balance were shown.

The theory of second-order circulants is well developed, and it was shown in [29] that circulants with generatrices, calculated by formula:

$$C(n; D-1, D), \text{ where } D = \sqrt{n/2}, \; n > 2, \tag{1}$$

were the optimal ones.

We also developed the software making it possible to synthesize circulants with specified characteristics [30] including ring circulants, optimal circulants, and circulants calculated by formula (1).

Thus, by synthesizing them, we are able to compare their characteristics with the characteristics of the mesh and torus graphs. Since the diameter and average distance between the nodes are one of the most important characteristics of NoC, we give the obtained characteristics for topologies with number of vertices from $3^2 = 9$ to $2^{32} =$ 529 (we take topologies only for the number of vertices which are numbers in the second degree; for mesh and torus topologies, the square form is the most optimal [14]). Dependencies of $L_{av}$ and $D$ on the number of vertices are shown in the figures below.

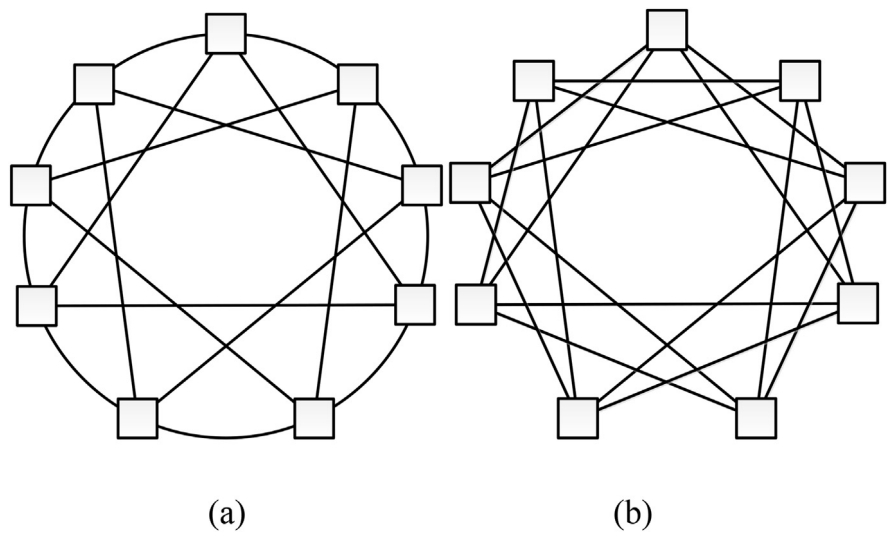

**Fig. 1.** Circulant topology: (a) — $C(9; 1, 3)$; (b) — $C(9; 2, 3)$.

Based on the obtained graphs, it can be concluded that circulants, at equal costs of connective resources as torus topology, make it possible to reduce: the diameter to 20.0–59.4 % in comparison with torus, and to 50.0–63.9 % in comparison with mesh; the average distance — to 2.1–6.7 % in comparison with torus, and to 15.6–29.2 % — in comparison with mesh. This is generally obvious, since torus can be represented as a circulant-like graph, but not the optimal one. In addition, in comparison with mesh and torus topologies, it is possible to construct circulant graphs for any number of vertices without reducing their effectiveness. Thus, the use of circulant graphs as a topological basis in NoC development makes it possible to improve NoC characteristics in comparison with mesh and torus topologies without losing the regularity of networks and increasing the dimension of routers. Moreover, like in other regular topologies, circulants interconnects can be reconfigured according to the application task graph by adding and removing some links, as proposed in work [31].

It should be noted that for the routing task, optimal circulants $C(n; D-1, D)$ may be worse than ring circulants, since the presence of a one-length generatrix simplifies routing algorithms in them. Nevertheless, as can be seen from the graphs (Fig. 2a), characteristics of circulants basically coincide, and differ insignificantly in $L_{av}$ only at some points (Fig. 2b). The difference in $D$ by 1 can be seen in Fig. 3 in the form of not frequent spikes. Thus, for some tasks, where it is required to simplify the routing algorithm, it is justified to use ring circulants, while they also provide a significant gain in their characteristics in comparison with the classical regular mesh and torus topologies.

## 4. Design

### 4.1. Development of the routing algorithms for NoCs with ring circulant topologies

#### 4.1.1. Table routing algorithm

The number of ports of the router is determined by the order of the circulant as $p = 2 \cdot k$, where $k$ — order of the graph [29]. Thus, a router of the ring circulant $C(N; 1, s_2)$ has 4 connections to other routers. To navigate in such a network, one can use the Table routing algorithm based on pre-calculated routes between routers in the network which are stored in the table. To operate the algorithm, the head flit should contain the destination node number. The load on the packet (the size of the address field, bit) can be calculated by the following formula:

$$P = \lceil \log_2 N \rceil, \tag{2}$$

where $N$ — number of nodes in the network.

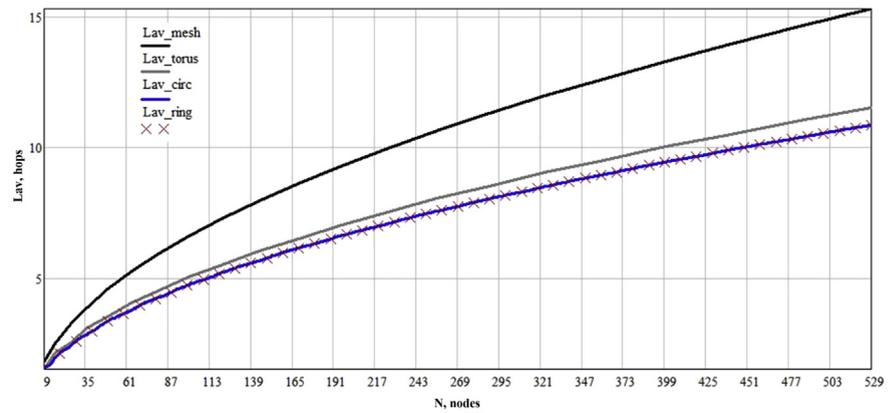

(a)

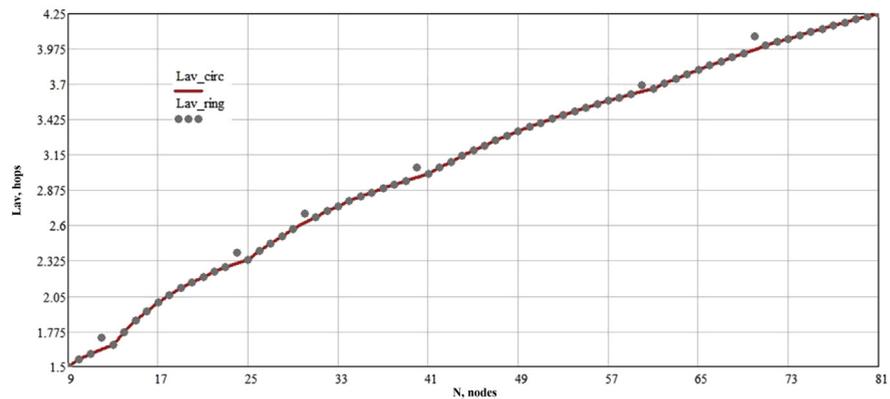

(b)

**Fig. 2.** Dependence of average distance between the shortest paths among all the nodes on the number of nodes: (a) – difference between circulant and mesh-like topologies; (b) – difference between optimal circulants and ring topologies.

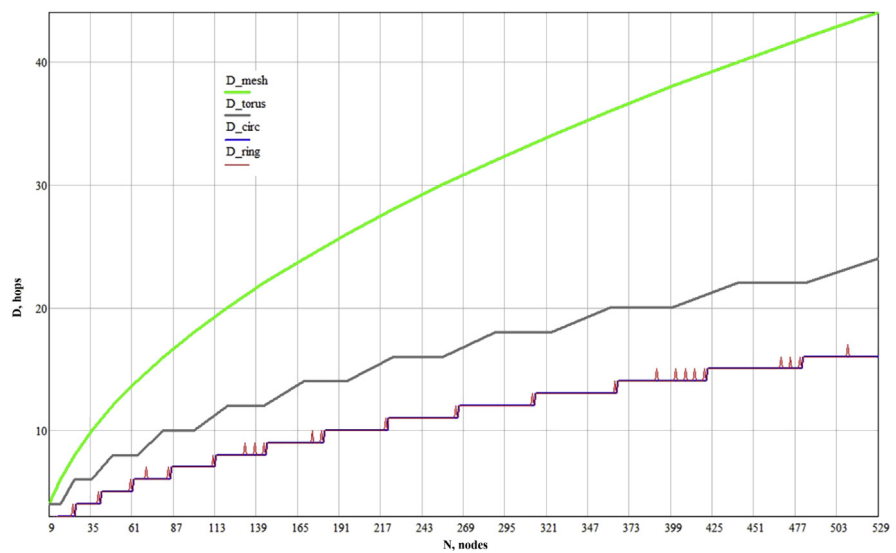

**Fig. 3.** Dependence of the diameter on the number of nodes.

The structure of the routing table is generally presented below (Table 1). The routers, being data sources, are given across, and the routers, serving as data receivers, are given vertically. At the intersection of every line and column, it is put the number of router output (in the direction of the clockwise movement from 0 to 3), where the packet should be sent to.

In Fig. 4, the routers are shown as numbered vertices of the graph, and bidirectional connections between them as edges of the graph. Numbers from 0 to 3 are port numbers of routers. This means that packets from the $0^{th}$ router to the $1^{st}$ router are sent via the $0^{th}$ port and received via the $3^{rd}$ port.

Thus, for router $C(8; 1, 3)$, the routing table will look like the one shown in Table 2. It should be said that such a table is not an incidence table, and therefore, is not symmetrical. It should be also noted that the routing table is stored in a network in a distributed manner. This is ensured by each router containing its own part of the table, which is the corresponding line, responsible for sending packets from the current router only.

The amount of memory, consumed by such a table, is described by the formula:

$$M = N^2 * \lceil \log_2 p \rceil \tag{3}$$

where $N$ — number of nodes in the network;

$\lceil \log_2 p \rceil$ — required amount of memory (bit) to store the port numbers of router;

$p$ — number of router ports.

The table routing algorithm is very simple and can be implemented as a multiplexer, which switches the packet to the port whose number corresponds to the destination node and is selected from the routing table. The use of this algorithm leads to an increased memory consumption, but at the same time, the state machine, implementing routing algorithm, is relatively simple and takes less logical resources of the chip.

**Table 1.** Routing table.

| From/To | 0 | 1 | 2 | … |
|---|---|---|---|---|
| 0 | - | 0 | 1 | … |
| 1 | 3 | - | 0 | … |
| 2 | 2 | 3 | - | … |
| … | … | … | … | - |

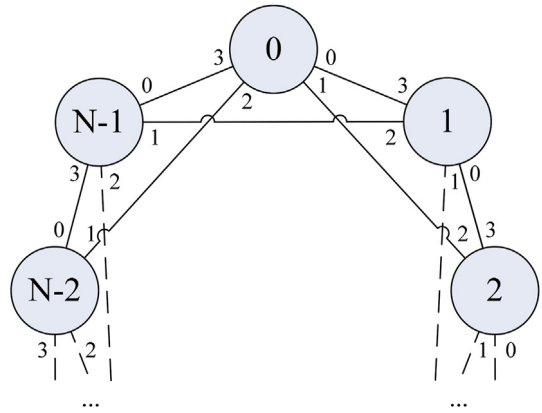

**Fig. 4.** Numbering of outputs of the router.

**Table 2.** Routing table for circulant $C(8; 1, 3)$

| From/To | 0 | 1 | 2 | 3 | 4 | 5 | 6 | 7 |
|---|---|---|---|---|---|---|---|---|
| 0 | — | 0 | 0 | 1 | 0 | 3 | 0 | 2 |
| 1 | 2 | — | 0 | 0 | 1 | 0 | 3 | 0 |
| 2 | 0 | 2 | — | 0 | 0 | 1 | 0 | 3 |
| 3 | 3 | 0 | 2 | — | 0 | 0 | 1 | 0 |
| 4 | 0 | 3 | 0 | 2 | — | 0 | 0 | 1 |
| 5 | 1 | 0 | 3 | 0 | 2 | — | 0 | 0 |
| 6 | 0 | 1 | 0 | 3 | 0 | 2 | — | 0 |
| 7 | 0 | 0 | 1 | 0 | 3 | 0 | 2 | — |

## 4.2. Clockwise routing algorithm

For navigation in ring circulants such as $C(N; 1, s_2)$, it is possible to use an algorithm based on iterative calculation of the route from the packet source router to the router receiver when each router decides to switch the packet to the next router by only one step. In the packet the difference between the nodes (source and receiver) is transmitted. The load on the packet is similar to the algorithm with table routing and is calculated by the formula (2). Additionally, in the router, it is necessary to store the number of nodes and the value of generatrix $s_2$. Thus, the total size of the stored data can be calculated using the following formula:

$$M = N * \left( \lceil \log_2 N \rceil + \left\lceil \log_2 \frac{N}{2} \right\rceil \right) \qquad (4)$$

where $N$ — number of nodes in the network;

$\left\lceil \log_2 \frac{N}{2} \right\rceil$ — required amount of memory (bit) to store the number of routers in the network;

$\left\lceil \log_2 \frac{N}{2} \right\rceil$ — required amount of memory (bit) to store generatrix $s_2$.

To store the value of generatrix $s_2$, it is necessary exactly $\left\lceil \log_2 \frac{N}{2} \right\rceil$ bits, since it is guaranteed to be less than half of the number of nodes [32]. Calculation of the transition is as follows: first, the direction of transition is determined (in the direction, or opposite to the direction of the clockwise movement); then the generator is selected. If the difference between the source and receiver nodes is less than half of the number of nodes, then the movement in the direction of the clockwise movement is chosen; if more — then the opposite direction is chosen. When the direction of the clockwise movement is chosen, the choice of the generator is as follows: until the difference between the source and receiver nodes is greater than the value of $s_2$, the transition will occur over a larger generatrix, otherwise — over the small generatrix. The value of address field of the head flit is recalculated by subtracting the length of the generatrix over which the transition will be made. Zero value of address field of the head flit is the criterion for the end of the packet transmission. If the movement was selected clockwise, the general algorithm for selecting the current step is preserved, but the comparison takes place between the larger generatrix and the difference in the number of nodes in the network and the value stored in the address field of the head flit. Before the transition, to the address value in the head flit, the value of the length of the generatrix, over which the transition occurs, is added. In this case, the criterion for the end of packet transmission is equality of the value of address field of the head flit to the number of nodes in the network. An algorithmic description of the proposed algorithm is given below:

**algorithm Find_Route_Clockwise is**

**Input:** $startNode$ — start node, $endNode$ — end node, $N$ — count of nodes, $s_1$ — first generatrix, $s_2$— second generatrix.

**Output:** $startNode$ — next start node.

1:   $S \leftarrow endNode - startNode$
2:  **if** $S = 0$ **then**
3:     **return** $startNode$
4:  **if** $S < 0$ **then**
5:     $S \leftarrow S + N$
6:  **if** $S \leq \frac{N}{2}$ **then**
7:     **if** $S \geq s_2$ **then**
8:        $startNode \leftarrow (s_2 + startNode) \bmod N$
9:     **else**
10:       $startNode \leftarrow (s_1 + startNode) \bmod N$
11:  **else**
12:     $S \leftarrow N - S$

13:    **if** $S \geq s_2$ **then**

14:        $startNode \leftarrow (N - s_2 + startNode) \bmod N$

15:    **else**

16:        $startNode \leftarrow (N - s_1 + startNode) \bmod N$

17:  **if** $startNode = 0$ **then**

18:    $startNode \leftarrow N$

19:  **return** $startNode$

It should be noted that this algorithm will also work with circulants of $C(N; 1, s_1, ..., s_n)$. To do this, it is enough to check additionally the possibility of moving along the generatrices in descending order of their index. Nevertheless, the presented algorithm is not optimal, because in some cases, it will offer paths that are larger than network diameter, but instead, it will significantly save the router's memory.

### 4.3. Adaptive routing algorithm

Evolution of the previous algorithm is an Adaptive algorithm that can change the direction of movement when calculating the transition to the next node. In this algorithm, it is proposed to store the destination node as an address in the head flit. The load on the packet remains the same as in the previous algorithm (2). In addition, the router must store its number, the number of nodes in the network, and the length of generatrix $s_2$. The total size of the stored data is calculated by the formula:

$$M = N * \left(2 * \lceil \log_2 N \rceil + \left\lceil \log_2 \frac{N}{2} \right\rceil \right) \tag{5}$$

where $N$ — number of nodes in the network;

$\left\lceil \log_2 \frac{N}{2} \right\rceil$ — required amount of memory (bit) to store the number of router and number of routers in the network;

$\left\lceil \log_2 \frac{N}{2} \right\rceil$ — required amount of memory (bit) to store generatrix $s_2$.

The operation of algorithm can be divided into two parts. In the first part, there is a choice of the sequence of transmission of the source and receiver nodes of the packet transmitted to the second part of the algorithm. This is possible because of the properties of circulant graphs. This procedure is required to simplify the algorithm for calculating the direction of movement due to the fact that it will work only with positive numbers. Also, in the first part, normalization of the chosen direction of the package movement occurs. In the second part of the algorithm, the next step of the packet's movement is calculated directly.

 

**algorithm Find_Route_Adaptive is**

**Input:** $startNode$ — start node, $endNode$ — end node, $N$ — count of nodes, $s_1$ — first generatrix, $s_2$ — second generatrix.

**Output:** $startNode$ — next start node.

1: **if** $startNode > endNode$ **then**
2:     $startNode \leftarrow startNode -$ **Step_Cycles**$(endNode, startNode, N, s_1, s_2)$
3: **else**
4:     $startNode \leftarrow startNode +$ **Step_Cycles**$(startNode, endNode, N, s_1, s_2)$
5: **if** $startNode > N$ **then**
6:     $startNode \leftarrow startNode - N$
7: **else**
8:     **if** $startNode \leq 0$ **then**
9:         $startNode \leftarrow startNode + N$
10: **return** $startNode$

**function Step_Cycles is**

**Input:** $startNode$ — start node, $endNode$ — end node, $N$ — count of nodes, $s_1$ — first generatrix, $s_2$ — second generatrix.

**Output:** the function returns the best step (direction is also selected)

1: $bestWayR \leftarrow 0$, $stepR \leftarrow 0$, $bestWayL \leftarrow 0$, $S \leftarrow endNode - startNode$
2: $R_1 \leftarrow \frac{S}{s_2} + S \bmod s_2$, $R_2 \leftarrow \frac{S}{s_2} - S \bmod s_2 + s_2 + 1$
3: **if** $S \bmod s_2 = 0$ **then**
4:     $bestWayR \leftarrow R_1$, $stepR \leftarrow s_2$
5: **else**
6:     **if** $R_1 < R_2$ **then**
7:         $bestWayR \leftarrow R_1$, $stepR \leftarrow s_1$
8:     **else**
9:         $bestWayR \leftarrow R_2$, $stepR \leftarrow s_2$
10: $R_5 \leftarrow \frac{S+N}{s_2} + (S+N) \bmod s_2$, $R_6 \leftarrow \frac{S+N}{s_2} - (S+N) \bmod s_2 + s_2 + 1$
11: **if** $R_5 <$ bestWayR **then**
12:     $bestWayR \leftarrow R_5$, $stepR \leftarrow s_2$
13: **if** $R_6 <$ bestWayR **then**
14:     $bestWayR \leftarrow R_6$, $stepR \leftarrow s_2$

15: $R_9 \leftarrow \frac{S+N+N}{s_2} + (S+N+N) \bmod s_2$, $R_{10} \leftarrow \frac{S+N+N}{s_2} - (S+N+N) \bmod s_2 + s_2 + 1$

16: **if** $R_9 <$ bestWayR **then**

17:     $bestWayR \leftarrow R_9$, $stepR \leftarrow s_2$

18: **if** $R_{10} <$ bestWayR **then**

19:     $bestWayR \leftarrow R_{10}$, $stepR \leftarrow s_2$

20: $S \leftarrow endNode - startNode + N$, $L_1 \leftarrow \frac{S}{s_2} + S \bmod s_2$, $L_2 \leftarrow \frac{S}{s_2} - S \bmod s_2 + s_2 + 1$

21: **if** $S \bmod s_2 = 0$ **then**

22:     $bestWayL \leftarrow L_1$, $stepL \leftarrow -s_2$

23: **else**

24:     **if** $L_1 < L_2$ **then**

25:         $bestWayL \leftarrow L_1$, $stepL \leftarrow (-s_1)$

26:     **else**

27:         $bestWayL \leftarrow L_2$, $stepL \leftarrow -s_2$

28: $R_7 \leftarrow \frac{S+N}{s_2} + (S+N) \bmod s_2$, $R_8 \leftarrow \frac{S+N}{s_2} - (S+N) \bmod s_2 + s_2 + 1$

29: **if** $R_7 <$ bestWayL **then**

30:     $bestWayL \leftarrow R_7$, $stepL \leftarrow -s_2$

31: **if** $R_8 <$ bestWayL **then**

32:     $bestWayL \leftarrow R_8$, $stepL \leftarrow -s_2$

33: $R_{11} \leftarrow \frac{S+N+N}{s_2} + (S+N+N) \bmod s_2$, $R_{12} \leftarrow \frac{S+N+N}{s_2} - (S+N+N) \bmod s_2 + s_2 + 1$

34: **if** $R_{11} <$ bestWayL **then**

35:     $bestWayL \leftarrow R_{11}$, $stepL \leftarrow -s_2$

36: **if** $R_{12} <$ bestWayL **then**

37:     $bestWayL \leftarrow R_{12}$, $stepL \leftarrow -s_2$

38: **if** $bestWayR < bestWayL$ **then**

39:     **return** $stepR$

40: **else**

41:     **return** $stepL$

This algorithm allows finding routes in circulants of $C(N; 1, s_2)$ in which the number of cycles does not exceed two. A cycle here is understood as such a route from node A to node B in which there is a transition through the reference point of the cycle (Fig. 5). For example, in circulant $C(100; 1, 44)$ the route from node 1 to node 38 (full path: 1−57−13−69−25−81−37−38) contains two cycles: 1−57−13−69 and 69−25−81−37. To eliminate them, there are lines 15−19 and

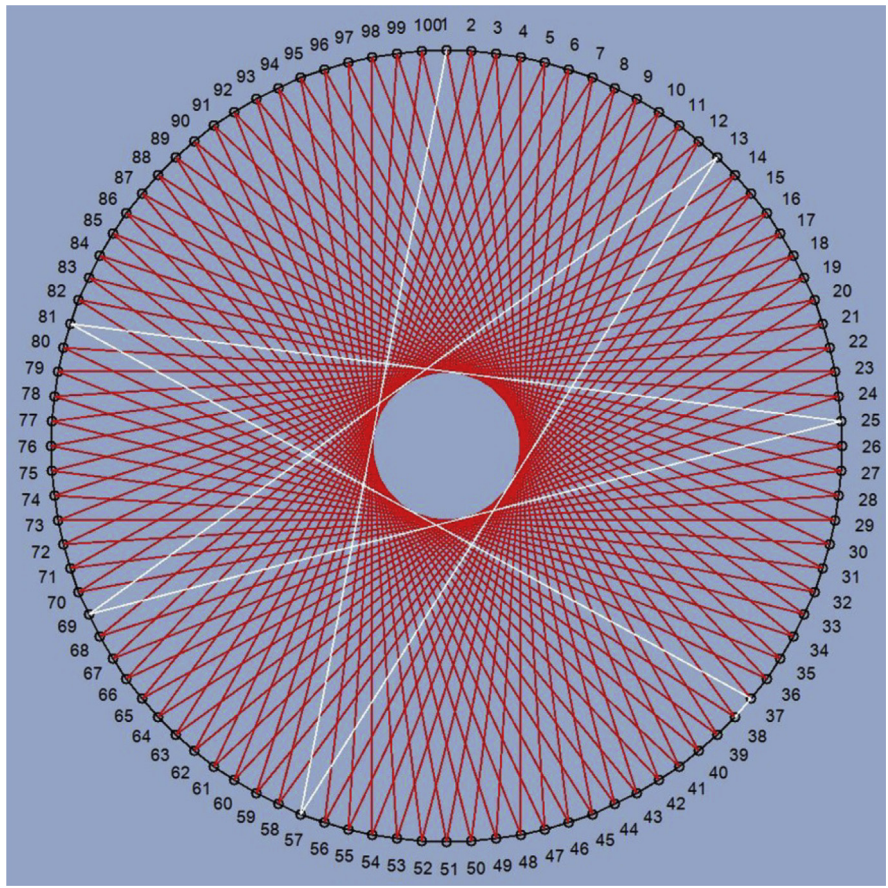

**Fig. 5.** Example of a route with two cycles in the circulant $C(100; 1, 44)$.

33–37 in **Step_Cycles** algorithm. The trend of changing the number of cycles in circulants depending on the number of nodes is shown in Fig. 6 (full data are provided in Appendix 1). Starting from $N = 174$, there are circulants, the number of cycles in which exceeds two. For such circulants, additional code lines should be added in a similar way. It should be noted that the proposed algorithm is specifically shown in expanded form without using loop operators to show how it becomes more complex with an increase in the number of cycles, and how the costs of combinational logic

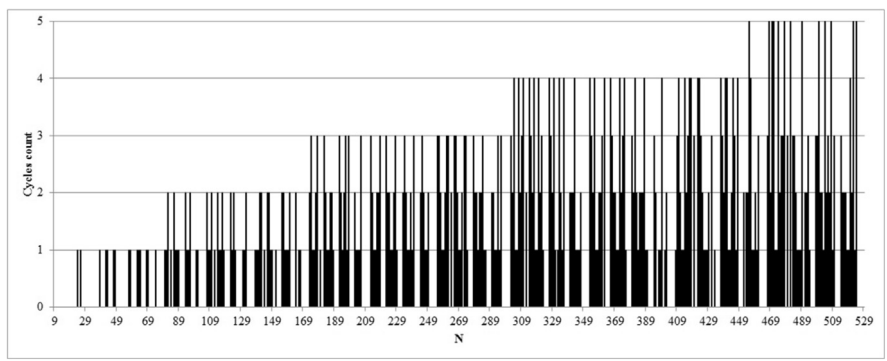

**Fig. 6.** Dependence of cycles count on network size.

 

for its implementation increase accordingly. This is done because this algorithm is designed for synthesis in the hardware description language as a combinational block, and this presentation is most easily shifted to what it will look like at the register transfer level after synthesis. To save logic, the same type of code lines at the hardware level can be replaced with a sequential digital automaton, but then the calculation of the route will take several cycles which will slow down the work of the whole circuit.

## 5. Calculation

### 5.1. Criterion of the effectiveness of algorithms

To evaluate the operation of algorithms, an efficiency criterion, taking into account the number of hops in which a packet could reach, from the first node, all the other ones, can be offered. As an optimal criterion, the Dijkstra algorithm [33], which guarantees the shortest paths between all the nodes of the network, can be used. Thus, the efficiency criterion is defined by the formula:

$$K = \frac{\sum_{i=1}^{N} HA_{(0-i)}}{\sum_{i=1}^{N} HD_{(0-i)}} \qquad (6)$$

where $N-$ number of nodes in the network;

$\sum_{i=1}^{N} HA_{(0-i)}$ — number of hops (from the 0 node to all the other ones) calculated by the algorithm used;

$\sum_{i=1}^{N} HD_{(0-i)}$ — number of hops (from the 0 node to all the other ones) calculated by the Dijkstra algorithm.

According to formula (6), the efficiency of the table routing algorithm will always be 1, because routes are calculated in advance and are optimal ones. For Clockwise algorithm, the efficiency is strongly dependent on the larger generatrix: the larger it is, the less is the efficiency of the algorithm (Fig. 7). For the adaptive algorithm on the circulants examined, the efficiency also turned out to be equal to 1, but for circulants with the number of cycles in routes greater than 2 (Fig. 6), it will decrease. To avoid this, a modification of the algorithm is required, but it will increase the consumption of logical resources on the routers.

## 6. Experimental

### 6.1. Approbation of algorithms operation

Testing of developed algorithms operation was carried out on FPGA Cyclone V 5CGXFC9A6U19I7 of Intel FPGA (Altera). The considered routing algorithms were implemented as separate modules in the Verilog language which makes it easy to integrate them into the description of any router. Testing was performed on the two routers: based on a router from the Netmaker [28] library implementing all the main technical solutions used in NoCs (wormhole routing, QoS arbitration, etc.), and on simplified routers from the NoCSimp library [34]. In addition, the modules, generating connections between routers to support the synthesis of circulant topologies, were modified. Since routing modules are implemented by separate blocks, their contribution to the consumption of crystal resources does not depend on the router itself. Therefore, the results of experiments and synthesis, carried out further, are based on the simpler NoCSimp library which allows placing more NoC nodes on limited crystal resources, but the same results are reproducible for more complex routers as well.

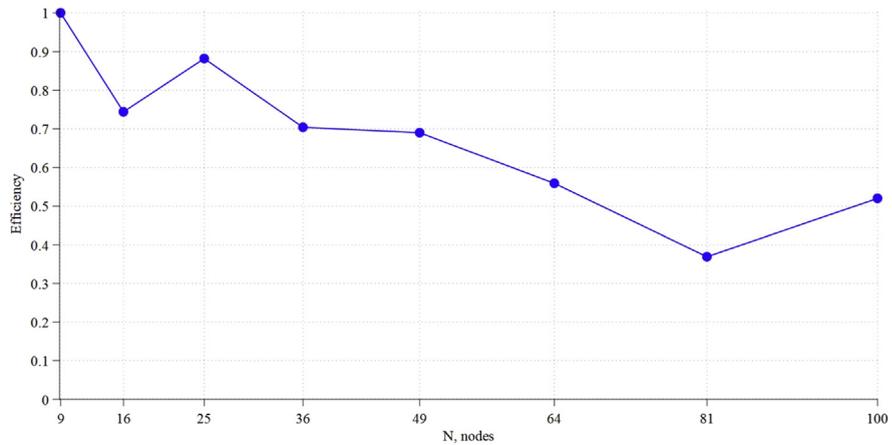

Fig. 7. Efficiency of Clockwise routing algorithm.

For testing, optimal circulants of $C(N; 1, s_2)$ with the number of nodes, determined by the formula $N = n^2$, where $N$ — number of nodes in the network, $n$ — natural number, were chosen. This is done in order to be able to compare the results of networks with circulant topologies with mesh and torus topologies for which the square shape of the topology is the optimal one (in order to eliminate the influence of the geometric form of topologies on comparison results) [14, 27].

Based on the above formulas (3–5), the values of the dependencies of memory, necessary for algorithm operation (bits), on the network size (Fig. 8), were theoretically obtained.

As a result of RTL synthesis, there was obtained the data on the number of ALM blocks consumed (Fig. 9) and registers (Fig. 10) on the FPGA.

With this compilation, the compiler option, implementing memory on registers, was chosen. This is not the most effective way, because the same ALMs are spent, but for the synthesis of memory. On the chip, there are specialized blocks of RAM, designed to store memory.

 

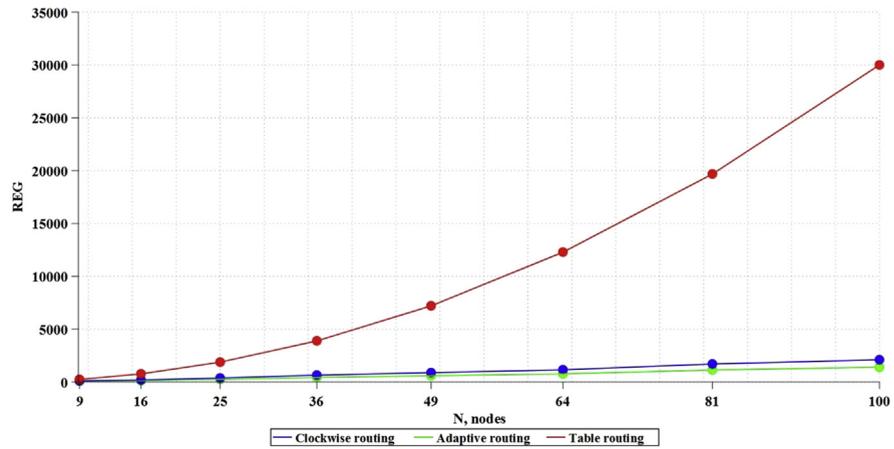

Fig. 8. Dependence of memory, consumed by algorithms, on network size.

Based on the obtained data, it is possible to derive formulas for theoretical calculation of the chip resources used before carrying out the simulation (full data and calculations are provided in Appendix 2). For Table routing, the formula $U_{almT}$ for calculating the ALMs used and the formula $U_{regT}$ for calculating the registers used will be:

$$U_{almT} = -74,354 + 15,537*x + 0,464*x^2; \qquad (7)$$

$$U_{regT} = 1163,150 - 9,069*x + 2,940*x^2. \qquad (8)$$

For the Clockwise algorithm, the formula $U_{almC}$ for calculating the ALMs used and the formula $U_{regC}$ for calculating the registers used will be:

$$U_{almC} = -93,577 + 22,553*x + 0,434*x^2; \qquad (9)$$

$$U_{regC} = -43,664 + 21,039*x + 0,270*x^2. \qquad (10)$$

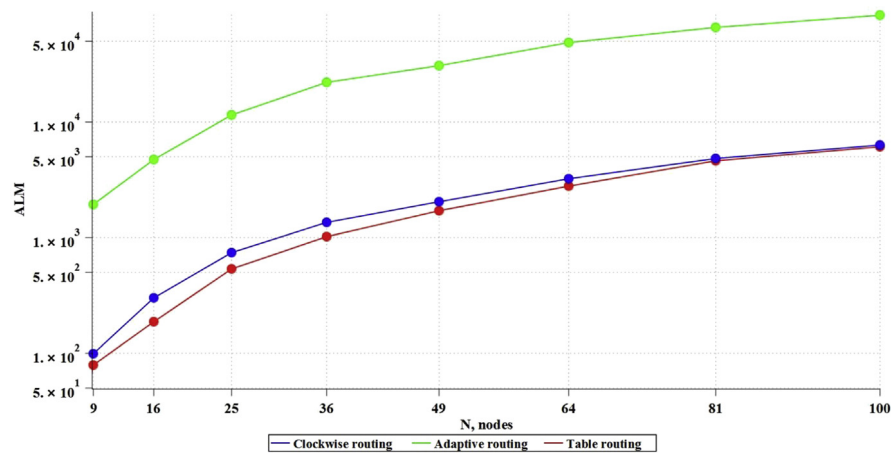

Fig. 9. Dependence of ALM blocks, consumed by algorithms, on network size.

 

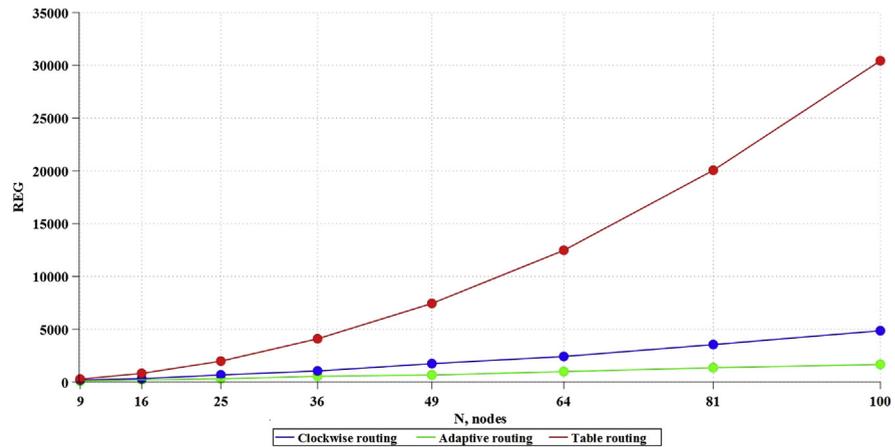

**Fig. 10.** Dependence of the registers, consumed by algorithms, on network size.

For the Adaptive routing algorithm, the formula $U_{almA}$ for calculating the ALMs used and the formula $U_{regA}$ for calculating the registers used will be:

$$U_{almA} = -6237,760 + 684,297*x + 3,329*x^2; \qquad (11)$$

$$U_{regA} = 1163,150 - 9,069*x + 2,940*x^2. \qquad (12)$$

Based on the fact that the SoC communication subsystem consumes 30−40 % of the total amount of chip resources [35], it is possible to estimate what size of NoC can fit in the chip, since one of the main limiting factors is the amount of ALMs and the number of RAM memory registers. Thus, on the chip 5CGXFC9A6U19I7, including 113560 ALM blocks and 12492800 RAM registers, in accordance with formulas (7−12), it is possible to place 275 routers with Table routing algorithm, 278 routers with Clockwise algorithm, and up to 53 routers with Adaptive routing algorithm. The work did not consider the speed of the network, and DSP blocks were not used in the synthesis; it is an expensive resource that will necessarily be used to implement NoC computational nodes. Thus, this work can be referred to those developing NoC communication subsystem.

## 7. Conclusion

Thus, we showed that classical mesh and torus topologies, used for NoC design, no longer satisfied modern requirements. There is a need for searching the topologies that will provide better characteristics in diameter and nodes average distance. One alternative is to use circulant topologies. We showed that ring circulant topologies had slightly worse characteristics than optimal circulant topologies, but they were much better than those possessed by mesh and torus.

The problem of applying new communication subsystem topologies for NoC is the development of routing algorithms in such topologies. We examined various routing strategies in such networks: Table routing, Clockwise routing and Adaptive routing. Table routing requires simple routers, which do not consume logical resources, but leads to increased consumption of memory resources. Route calculation is carried out in advance and is static. In algorithms of Clockwise routing and Adaptive routing, each router decides on how to route a packet to a specific port on the path of the packet. Clockwise routing is not the optimal one: because of simplifying the algorithm, the path of the packet, in contrast to Adaptive routing, is not the shortest. Complication of routing algorithm (Adaptive routing) entails an increase in the costs of logical resources of the chip.

Synthesis of communication subsystems for NoCs of different sizes for different types of routing was carried out and formulas for estimating the costs of chip resources, depending on the number of nodes, were calculated. This allowed us to estimate how many routers can theoretically fit on an average static FPGA chip. As a result, it was shown that ALMs were a more critical resource than RAM which limited the number of routers. Thus, Table routing, or significantly simplified Clockwise routing is preferred. Nevertheless, it is required to check the speed of algorithms, as well as to simulate NoCs with such communication subsystems.

## Declarations

### Author contribution statement

Aleksandr Yu. Romanov: Conceived and designed the experiments; Performed the experiments; Analyzed and interpreted the data; Wrote the paper.

### Funding statement

The research leading to these results has received funding from the Basic Research Program at the National Research University Higher School of Economics.

### Competing interest statement

The authors declare no conflict of interest.

### Additional information

Data associated with this study has been deposited at https://github.com/RomeoMe5/routingAlgorithms, https://github.com/RomeoMe5/NoC_HDL_model/tree/master/NoCSimp, and https://github.com/RomeoMe5/circulantGraphs.

Supplementary content related to this article has been published online at https://doi.org/10.1016/j.heliyon.2019.e01516.

## Acknowledgements

We acknowledge A.A. Amerikanov and E.V. Lezhnev from National Research University Higher School of Economics for technical assistance and some calculations provided.